\documentclass[12pt]{iopart}

\usepackage{graphicx}
\begin{document}

\title{RE L$_{3}$ X-ray absorption study of REO$_{1-x}$F$_{x}$FeAs (RE=La, 
Pr, Nd, Sm) oxypnictides}

\author{B. Joseph$^{1}$, A. Iadecola$^{1}$, M. Fratini$^{2}$, A.
Bianconi$^{1}$, A. Marcelli$^{3}$, N.L. Saini$^{1}$}
\address{$^{1}$Dipartimento di Fisica, Universit\`{a} di Roma ``La 
Sapienza", P. le Aldo Moro 2, 00185 Roma, Italy}
\address{$^{2}$Istituto di Fotonica e Nanotecnologie, CNR Roma, Italy}
\address{$^{3}$Laboratori Nazionali di Frascati, INFN, 00044 Frascati, 
Italy}

\begin{abstract}

Rare-earth L$_{3}$-edge X-ray absorption near edge structure (XANES)
spectroscopy has been used to study REOFeAs (RE=La, Pr, Nd, Sm)
oxypnictides.  The Nd L$_{3}$ XANES due to 2p$_{3/2}\to$5$\epsilon$d
transition shows a substantial change in both white line (WL) spectral
weight and the higher energy multiple scattering resonances with the
partial substitution of O by F. A systematic change in the XANES
features is seen due to varying lattice parameters with ionic radius
of the rare earth.  On the other hand, we hardly see any change across
the structural phase transition.  The results provide timely
information on the local atomic correlations showing importance of
local structural chemistry of the REO spacer layer and interlayer coupling
in the competing superconductivity and itinerant striped magnetic
phase of the oxypnictides.

\end{abstract}

\pacs{61.05 Cj;74.62 Bf;74.81 -g}

\maketitle

\section{Introduction}

Recently, the discovery of high temperature superconductivity in the
Fe-based oxypnictides \cite{Kamihara} has attracted wide attention
from the condensed-matter community.  The rare-earth (RE) oxypnictides
(REOFeAs) have a layered structure, similar to the copper oxide
superconductors and diborides, with electronically active
(FeAs)$^{\delta-}$ layers separated by the (REO)$^{\delta+}$
reservoirs.  Both the charge transfer and the lattice misfit between
the two sub-layers \cite{MFRicci,Fratini} are important for the
superconducting function of these materials similar to the case of
cuprates \cite{MFBianconi,MFPoccia} and diborides \cite{MFAgrestini}.
The REOFeAs get superconducting with maximum transition temperature
T$_{c}\approx$ 55 K \cite{RenEPL,HosRev} when the active layers are
doped through atomic substitutions or oxygen vacancies in the REO
spacers.  One of the interesting aspects of these materials is the
competing spin density wave (SDW) and superconductivity
\cite{HosRev,JZhaoNatP,HuangPRB,JZhaoPRB}.  Indeed, the undoped
compound REOFeAs is antiferromagnetically ordered (albeit a poor
metal), however at the same time showing a structural phase transition
from the tetragonal to the orthorhombic phase
\cite{DCruzNat,Fratini,Margadonna}.  With doping the system gets
superconducting and the structural transition as well as the SDW
transition disappear \cite{HosRev,JPSJIssue,NewJP,Izyumov}.  In
addition, while the maximum T$_{c}$ of the doped system increases with
reducing the RE ion size \cite{RenEPL,CHLee}, the structural
transition temperature (T$_{s}$) decreases for the undoped system
\cite{Fratini}.  This further underlines importance of local chemistry
of the REO layers in the correlating itinerant magnetism and
superconducting function of these materials.

It has been well known that RE L$_{3}$ X-ray absorption near-edge
structure (L$_{3}$-XANES) spectroscopy is a direct probe of the local
structure around a selected absorbing atom and distribution of the
valence electrons, with the final states in the continuum being due to
multiple scattering resonances of the photoelectron in a finite
cluster \cite{Konings}.  Moreover, in an ordered structure, the focusing effects of
atoms in collinear geometry determines multiple scattering resonances
also in a large cluster of near neighbor atoms.Therefore the RE
L$_{3}$-XANES is an ideal tool to study local geometry around the RE
and has often been applied to study the rare earth containing
materials \cite{Krill,Bianconi,Liu08}.  The added advantage of this
technique lies in the fact that, unlike the photoemission experiments,
there are negligible surface effects (and multiplet effects), making
it a very useful finger print probe of unoccupied valence states and
local chemistry of the absorbing atom.

Here we have exploited the RE L$_{3}$-XANES spectroscopy to make a
systematic study of the local geometry around the RE atom and the
related electronic structure, measuring a series of REOFeAs (RE=La,
Pr, Nd and Sm) with the varying rare earth size.  We have also studied
effect of tetragonal to orthorhombic structural phase transition as
well the superconductivity on the RE L$_{3}$ XANES. We find a
systematic change in the L$_{3}$-XANES spectra reflecting varying
electronic and local geometrical structure of the title system.
Incidentally the L$_{3}$-XANES white lines hardly show any change
across the tetragonal to orthorhombic structural phase transition
temperature, however, a clear change is observed with the electron
insertion by F-substitution.  The results suggest that, in addition to
the FeAs layers, structural coupling with the spacer layer should be
important for the high T$_{c}$, recalling the recent debate on misfit
strain \cite{MFPoccia,MFBianconi,MFAgrestini} and out of plane
disorder \cite{Eisaki,CSun,Hobou} controlling superconductivity
in the copper oxide perovskites.

\section{Experimental Details}

RE L$_{3}$-edge X-ray absorption near edge structure (L$_{3}$-XANES)
measurements were performed on powder samples of REOFeAs (RE=La, Pr,
Nd, Sm) prepared using solid state reaction method \cite{RenEPL2}. 
Prior to the absorption measurements, the samples were characterized
by x-ray diffraction for their structural properties \cite{Fratini}. 
The structural phase transition temperatures were measured to be 165
K, 155 K, 135 K and 130 K respectively for the La, Pr, Nd and Sm
containing REOFeAs samples.  The superconducting transition
temperature for the F-doped NdO$_{1-x}$F$_{x}$FeAs is found to be 49 K
 \cite{Digio}.  The X-ray absorption measurements were made at the XAFS
beamline of the Elettra Synchrotron Radiation Facility, Trieste, where
the synchrotron radiation emitted by a bending magnet source was
monochromatized using a double crystal Si(111) monochromator.  The
L$_{3}$-XANES measurements were made at two temperatures (room
temperature and 80 K) in the transmission mode using three ionization
chambers mounted in series for simultaneous measurements on the sample
and a reference.  For the low temperature measurements, the samples
were mounted on the liquid nitrogen cryostat cold finger enclosed in
an Al shroud with a Be window.  The sample temperature was measured to
be 80$\pm$1 K for the low temperature measurements.  As a routine
experimental approach, several absorption scans were collected to
ensure the reproducibility of the spectra, in addition to the high
signal to noise ratio.

\section{Results and Discussion}

Figure 1 shows normalized RE-L$_{3}$ spectra of REOFeAs (La, Pr, Nd,
Sm) measured at room temperature showing an intense peak, the
characteristic white line (WL) of RE$^{3+}$.  Here, the zero of the
energy scale is fixed to the WL peak position for a proper comparison
between the different spectra.  It is worth recalling that the L$_{3}$
absorption process is a \textit{2p$_{3/2}\to$5$\epsilon$d} (or
\textit{2p$_{3/2}\to$6$\epsilon$s}) transition governed by the dipole
selection rules (\textit{l} = $\pm$1) and hence empty states with
\textit{d} or \textit{s} symmetries (and admixed states) can be
reached in the final state.  Since the probability for a
\textit{2p$_{3/2}\to$6$\epsilon$s} transition is about two orders of
magnitude lower than for a \textit{2p$_{3/2}\to$5$\epsilon$d}
transition, the earlier can be ignored for describing the L$_{3}$ WL.
The one-electron picture (i.e., all the orbitals not directly involved
in the absorption process are passive in the final state) works well
for the L$_{3}$ WL unless or otherwise the materials of interest show
phenomena as the mixed valence.  In the present case, the WL appears
to be a typical of RE$^{3+}$ and hence the one electron picture could
be fairly used to describe the spectra.

In addition to the intense WL, we can see different near edge
features, a weak structure around 15 eV above the WL (feature A), and
the continuum resonances appearing as a two peak structure (the broad
features denoted by B$_{1}$ and B$_{2}$ appearing respectively around
35 eV and around 50 eV) due to the photoelectron scattering with the
nearest neighbors.  The RE atoms have 4 near neighbors O sitting at a
distance $\approx$ 2.3 $\AA$ and 4 As atoms occupying a site at
$\approx$ 3.3 $\AA$, and hence the two peak structure of the continuum
resonance is expected with the low and high energy peaks due to As and
O scatterings respectively.  The energy separation between the two is
consistent with the $\Delta$E$\propto$1/d$^{2}$ relation for a
continuum resonance \cite{BianconiXANES}.  Here it is worth referring Ignatov
et al \cite{Tyson} reporting the La L$_{3}$ -XANES on LaOFeAs with a
typical La$^{3+}$ WL and the continuum peak, however, with a limited
discussion they have incorrectly assigned the continuum resonance Peak
B$_{1}$ to the La-O distance.

The feature A is very weak or absent for the La and Pr L$_{3}$ -XANES
while clearly visible for the Nd and Sm L$_{3}$ cases (see e.g., the
inset).  The feature A has been frequently observed in the
L$_{3}$-XANES of the rare earths and discussed to be due to a
particular local structural geometry of the RE \cite{Tan}.  Here the
feature A is found to be absent (or very weak) in the La and Pr
L$_{3}$ spectra while clearly seen in the Nd and Sm L$_{3}$ edges.
The feature A is found to appear at the same energy.  Indeed, the
observation is consistent with the earlier \cite{Tan,Wu}, albeit with
dubious interpretation relating it to the local structure.
Considering the fact that, the basic geometry of the RE is the same,
it is difficult to rule out the feature A to have electronic origin or
be related with disorder in the REO layer, however, detailed multiple
scattering calculations should be performed for further clarification.

On the other hand, the continuum resonance Peak B$_{1}$ and Peak
B$_{2}$ show a systematic change with the rare earth ion size (see
e.g. the inset) indicating a clear evolution of local geometry around
the RE. The Peak B$_{1}$ and Peak B$_{2}$ shift towards higher energy
due to decreasing RE-As and RE-O distances with decreased rare earth
size, consistent with the $\Delta$E$\propto$1/d$^{2}$ empirical rule
\cite{BianconiXANES}.  It is interesting to note that, neither the WL
nor the XANES features show any appreciable change across the
tetragonal to orthorhombic structural transition temperature (dotted
and solid lines in Fig.  1).  Therefore, while the average symmetry
measured by diffraction (long range ordered structure) changes, local
atomic structure measured by the XANES hardly show any change except a
thermal shrinkage of the lattice (Figure 2).  On the other hand,
negligible change in the WL intensity suggests that electronic
structure of the unoccupied RE d-states (and admixed electronic
states) remains the same in the two structural phases.  This
observation is consistent with the Fe K-edge XANES \cite{Mustre} study
revealing hardly any change in the local structure across the
structural phase transition in these materials.  However, the Fe
K-edge study has revealed a clear change in the 1s$\to$3d quadrupole
transition (albeit small) across the structural phase transition.
Thus, it is possible that the tetragonal-to-orthorhombic structure
phase transition being induced by the electronic part due to the fact
that the electronic sates near the Fermi surface are mainly of Fe 3d
nature.  However, more work is necessary to discuss the structural
phase transition versus local structure by XANES, decorated by
polarized data.

Figure 2 shows energy positions of the Peak B$_{1}$ with respect to the
WL ($\Delta$E) at two temperatures, showing a continuous
increase with decreasing rare earth size due to decreased RE-As
distance.  The $\Delta$E at low temperature is slightly higher almost
by same amount, suggesting an average but similar unit cell shrinkage
for all the samples by lowering the temperature.  We have also plotted
the crystallographic RE-As distance at room temperature for a comparison
(lower panel).  The inset displays the $\Delta$E as a function
1/d$^{2}$, showing an expected linear behavior \cite{BianconiXANES}.

To get further information from the L$_{3}$-XANES, we have extracted
the line shape parameters for the WL as well as the Peak B$_{1}$ and
Peak B$_{2}$.  In the one-electron approximation the L$_{3}$-XANES WL
could be deconvoluted in a Lorentzian (core hole life time) convoluted
by the experimental broadening providing 5d (and admixed) density of
empty states weighted by the matrix element and the arctan curve to
take into account the transition into the continuum.  The energy
positions are similar to what have been obtained by second derivative
of the spectra (Fig.  2).  All efforts were made to take care of
experimental aspects influencing the L$_{3}$-line shapes (thickness
and homogeneity of the samples), however, it is not correct to
consider equal transition matrix elements for all rare earths,
therefore the absolute intensities can not be compared quantitatively
as a function of the RE. Moreover, the absolute intensity of the WL
does not show any appreciable or systematic change with the rare
earth.  On the other hand, the continuum resonance Peak B$_{1}$ and
Peak B$_{2}$ intensities, probing the local geometry around the rare
earth atom change systematically.  The relative change in the
intensity of the Peak B$_{1}$ and Peak B$_{2}$, defined as
\textit{$\Delta$I=[I(B$_{1}$)-I(B$_{2}$)]/[I(B$_{1}$)+I(B$_{2}$)]}, is
displayed in Figure 3 as a function of rare earth, showing a decrease
with decreasing rare-earth size, revealing a systematic change of
local geometry around the RE involving the As and O atoms (Peak
B$_{1}$ is due to RE-As while the Peak B$_{2}$ is due to RE-O
distance).  Since the RE-As represents out of plane atomic correlation
while the RE-O describes in-plane correlation, the decreasing
$\Delta$I should reflect increased correlation between the RE-O layer
and the Fe-As layer (i.e., increased interlayer coupling) with
decreasing rare earth size.  Although, it is difficult to quantify the
atomic displacement pattern around the RE, the results seems to be
consistent with the higher perpendicular position of the As atom with
respect to the Fe-Fe plane (Fig.  3b) and decreased RE-As distance
(Fig.  2b) with the decreasing rare earth size.

Let us tune to the effect of doping on the L$_{3}$-XANES. To explore
this, we have used NdOFeAs as a representative and measured Nd
L$_{3}$-XANES on the undoped and F-substituted system.  Figure 4 shows
the measured Nd L$_{3}$-XANES at room temperature on the NdOFeAs and
the superconducting NdO$_{1-x}$F$_{x}$FeAs system (T$_{c}\sim$49 K).
Both samples are in their tetragonal phase and hence the differences
reflect the effect of F-substitution.  The L$_{3}$-edge WL intensity
shows a significant increase with the F-doping, suggesting an increase
of a localization of the RE empty \textit{d} density of states
\cite{Chaboy}.  In addition, a substantial change in the Peak B$_{1}$
and Peak B$_{2}$ can be seen (inset of Fig 4).

Recently Sun et al \cite{Sun} have studied CeO$_{1-x}$F$_{x}$FeAs by
Ce L$_{3}$-XANES system as a function of external pressure revealing a
drastic decrease of the WL intensity while the T$_{c}$ decreases with
increasing hydrostatic pressure.  Incidentally, the results reported
here show more intense Nd L$_{3}$ WL for the superconducting system,
and hence increased intensity (more atomic like character) of the
L$_{3}$ WL may be somehow related with the superconductivity.  It is
possible that the F-substitution in the NdOFeAs causes reduced
hybridization between the Nd \textit{5d} and admixed \textit{4f}
orbitals, similar to the 4f-supercondutors \cite{HeavyFermion} in
which the superconductivity gets suppressed with increased
hybridization.

On the other hand, the Peak B$_{1}$ seems to show different effect of
F-doping.  Indeed, Peak B$_{1}$ and Peak B$_{2}$ tend to merge (Peak
B$_{1}$ hardly show any shift while the Peak B$_{2}$ appears at lower
energy) The doping of F-induces an increase of Nd-O/F distance and
hence the shift of the Peak B$_{2}$ towards lower energy is
reasonable.  On the other hand, the diffraction data reveal shrinkage
of the RE-As bonds, and hence a shift of the Peak B$_{1}$ towards
higher energy is expected, inconsistent with the present results.
This may be due to the fact that diffraction tend to average local
distortions and is less sensitive to localized structural distortions
and possible nanoscale inhomogeneities\cite{Oyanagi,Sato}.

In summary, we have studied RE L$_{3}$-XANES of the oxypnictides
REOFeAs with variable rare-earth ion to explore the valence electronic
structure as well as the local geometry of the RE-O slab.  While the
L$_{3}$ white lines hardly show any change across the tetragonal to
orthorhombic structural phase transition temperature, a clear change
is observed in the spectral weight with the electron doping.  In
addition, the XANES features above the WL show a systematic change due
to varying local geometry with the rare-earth size.  The energy
separation between the WL and the continuum resonances change
consistently with a change of RE-As and RE-O distance due to changing
misfit between the FeAs and REO sub-layers.  A systematic change of
the relative intensities of the continuum resonance features is also
observed.  Since the vertical position of the As atom change
systematically with the rare earth size (As atom moves away from the
Fe-Fe plane due to decreased Fe-As-Fe angle with decreasing rare earth
size \cite{Iadecola}), this has a direct influence on the local
geometry of the REO slabs, as demonstrated by the XANES spectra.  The
results suggest that the inter-layer coupling between the Fe-As and
the spacer (REO) layer may play a role also in the superconductivity
mechanism of the REOFeAs.  Thus, the outcome of this work recalls
recent debate on the misfit strain \cite{MFPoccia,MFBianconi} and out
of plane disorder controlled superconductivity in the copper oxide
high T$_{c}$ superconductors \cite{Eisaki,CSun,Hobou}, bringing
the two families on the same platform for the discussion.  Although,
it is difficult to argue on the cause or consequence of the local
structural changes, and quantitative role of intra and interlayer
coupling in the correlating itinerant magnetism and superconductivity,
the present results certainly add a new experimental feed-back to
understand the role of structure and local chemistry in fundamental
properties of REOFeAs oxypnictides.

\section*{Acknowledgments}

The authors thank the Elettra staff for the help and cooperation
during the experimental run (proposal no.  20085317).  We also
acknowledge Zhong-Xian Zhao and Z.A. Ren (Beijing) for providing high
quality samples for the present study.  This research has been
partially supported by the COMEPHS (FP6-STREP Controlling mesoscopic
phase separation).  One of us (BJ) would like to acknowledge the MIUR
(Italy) for a fellowship under the Italy-India bilateral programme.

\begin{figure}
\begin{center}

\includegraphics[width=120mm]{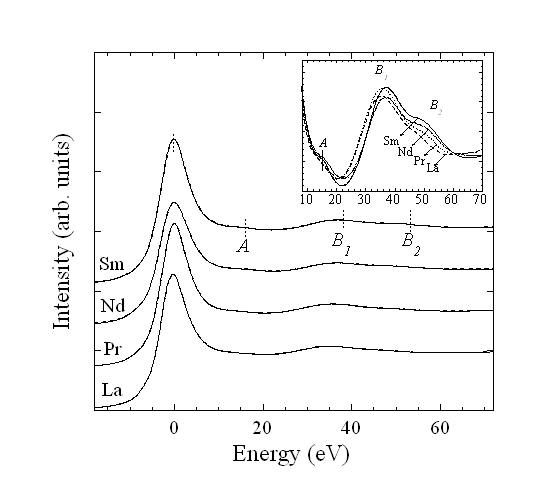}
\caption{\label{fig:epsart}The near-edge regions of the normalized RE
L$_{3}$-edge absorption spectra of the REOFeAs at 80 K and room
temperature (solid and dotted lines respectively).  The zero of the
energy scale is fixed to the characteristic WL representing
2p$_{3/2}\to$5d transitions.  Other near edge features are assigned as
A, B$_{1}$ and B$_{2}$, showing change of local geometry with changing
rare earth.  The inset shows a zoom over the near edge features A,
B$_{1}$ and B$_{2}$ (room temperature).}
\end{center}
\end{figure}

\begin{figure}
\begin{center}
\includegraphics[width=120mm]{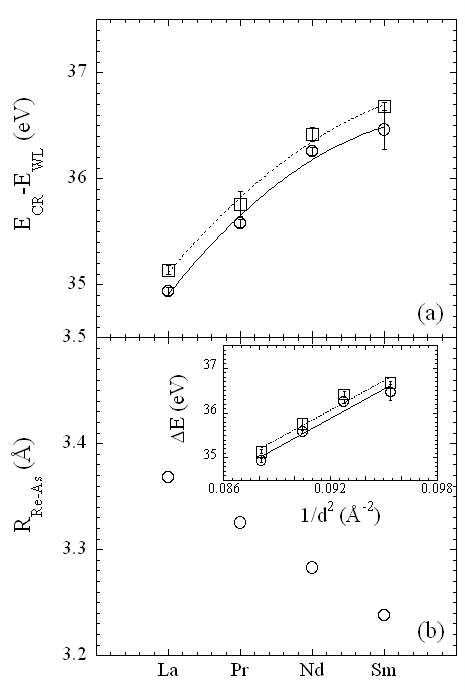}
\caption{\label{fig:epsart}Energy separation between the white line
and the continuum resonance Peak B$_{1}$ ($\Delta$E) at room
temperature (open circles) and at 80 K (open squares), showing a
continuous increase with decreasing rare earth size due to decreased
RE-As distance (a).  The energy positions are determined from second
derivative of the L$_{3}$ XANES. The crystallographic RE-As distance
at room temperature is also shown for comparison (b).  As expected, a
linear relation is observed between the $\Delta$E and 1/d$^{2}$ (see
inset in 2b).  The error bars represent standard deviation estimated
using different experimental scans.}
\end{center}
\end{figure}

\begin{figure}
\begin{center}
\includegraphics[width=120mm]{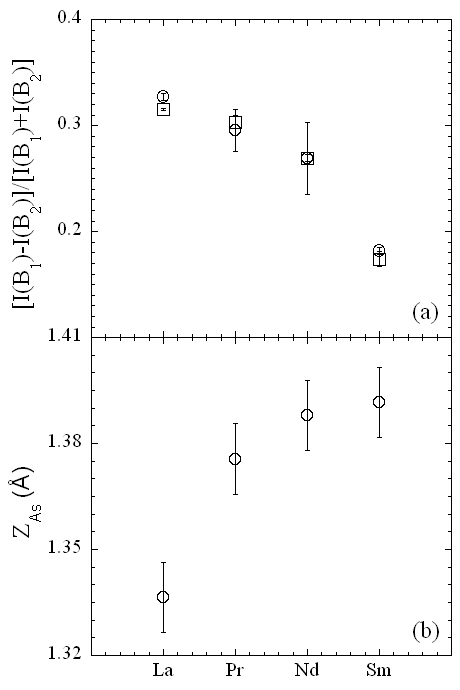}
\caption{\label{fig:epsart}Difference of relative intensities of the
continuum resonances B$_{1}$ and B$_{2}$ as a function of rare earth
(a).  The vertical position of the As atom with respect to the Fe-Fe
plane, measured by the Fe K-edge EXAFS \cite{Iadecola} is also shown
(b).}
\end{center}
\end{figure}

\begin{figure}
\begin{center}
\includegraphics[width=120mm]{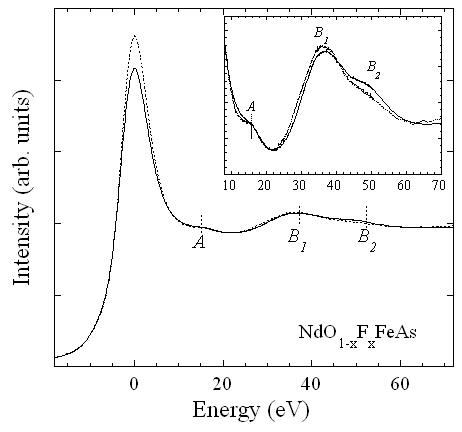}
\caption{\label{fig:epsart}Nd L$_{3}$-edge for the undoped (solid) and
F-substituted (dotted) superconducting systems at room temperature.
Both the samples are in their tetragonal phase and hence the
differences are merely due to the F-insertion.  Inset shows a zoom
over the features A, Peak B$_{1}$ and Peak B$_{2}$.}
\end{center}
\end{figure}

\end{document}